\documentclass[10pt]{article}

\usepackage{amsmath}
\usepackage{amssymb}

\usepackage{graphicx}

\usepackage{cite}

\usepackage{color} 

\usepackage[labelfont=bf,labelsep=period,justification=raggedright]{caption}

\date{}

\begin{document}

\begin{flushleft}
{\Large
\textbf{Analysis of Barcode sequence features to find anomalies due  to amplification Bias} 
\linebreak
}
\\
\bf Chandrima Sarkar, Raamesh Deshpande, Chad Myers  \linebreak
\\

\bf Department of Computer Science, University of Minnesota at Twin cities, Minneapolis, MN, USA
\\

\end{flushleft}

\section*{Abstract}

In this paper we aim at investigating whether barcode sequence features can predict the read count ambiguities caused during PCR based next generation sequencing techniques. The methodologies we used are mutual information based motif discovery and  Lasso regression technique using features generated from the barcode sequence. The results indicate that there is a certain degree of correlation between motifs discovered in the sequences and the read counts. Our main contribution in this paper is a thorough investigation of the barcode features that gave us useful information regarding the significance of the sequence features and the sequence containing the discovered motifs in prediction of read counts.


\section*{Introduction}

Next-generation DNA sequencing (NGS) technologies, such as the 454-FLX (Roche), SOLiD (Applied Biosystems) and Genome Analyzer (Illumina) have transformed the landscape of genetics with their ability to produce thousands of megabases of sequence information in a very short time  \cite{margulies2005genome}.  These kinds of sequencing techniques permit the rapid production of large and small sequence data sets. Among the popular strategies, Multiplexing technique is the one which allows parallel sequencing of many different samples \cite{hamady2008error, parameswaran2007pyrosequencing}. In this technology a unique sample specific fixed length marker sequence known as the 'barcode' is added to the DNA that is to be sequenced. Reads obtained after sequencing are sorted into sample libraries via detection of the appropriate barcode. Constructing mixtures of bar-coded or tagged DNA templates for sequencing can prove to be extremely useful in many applications \cite{binladen2007use, shokralla2012next}. However, though these technologies provide a great deal of benefits such as detection of nucleotide one at a time , accurate sequencing of homopolymer regions and many more, there are disadvantages too, such as short-read length causing a limitation in situations where no reference sequence is available to align, assign and annotate the short sequences generated. Another major drawback which all the PCR based NGS systems share is the amplification bias. It has been observed that barcode-dependent bias is highly prominent in DNA and RNA detection and expression profiling using multiplexing technology  \cite{alon2011barcoding,berry2011barcoded }. For production of high quality multiplex amplicon libraries for high-throughput sequencing it is therefore extremely important to detect the actual cause of this anomaly. A lot of studies have addressed this problem \cite{shiroguchi2012digital, aird2011analyzing, berry2011barcoded, casbon2011method, alon2011barcoding} from various dimensions. However, according to the best of our knowledge, none of the previous studies have tried to analyse the barcode sequence features which may prove to be the potential cause behind this existing bias. In this paper we try to analyse the characteristic features of the barcode sequences and classify them if any observable difference exists causing the differential behaviour during the PCR amplification. We consider the read counts as the expression profile directly indicating the growth rate for a given condition. 

To Summarize, The main aim of this paper can be given as -
\begin{itemize}
\item
Predict read counts using methods such as classification and prediction technique and mutual information based motif extraction.
\item
Predict  if Barcode sequence features are responsible for amplification biases in PCR based multiplexing techniques.
\end{itemize}

\subsection*{Background}

Due to the availability of whole genome sequences over the past decade, there has been major developments in the reverse genetics approaches. This technological surge has caused dramatic changes to the way gene functions are analysed. The ultimate reverse genetics tool, whole-genome deletion mutant libraries, were first created for the budding yeast Saccharomyces cerevisiae \cite{giaever2002functional, winzeler1999functional}. This tool helps in predicting the open reading frames in the budding yeast genome which can be used for analysing the phenotypes of their deletion mutants. New genes involved in various biological pathways can now be predicted with the help of buddying yeast deletion libraries \cite{scherens2004uses}. Another significant application of the deletion libraries is profiling drug sensitive yeast mutants for targeting the therapeutic compounds for analysing drug-gene interactions \cite{giaever2004chemogenomic}. The construction of the budding yeast deletion libraries incorporated the ingenious idea of molecular barcodes, for deletion cassette \cite{shoemaker1996quantitative}. With the help of these barcodes the mutants could be phenotyped by allowing thousands of mutant strains to be pooled and analyzed together in a highly parallel fashion. The barcodes can be easily amplified by PCR from genomic DNA extracted from the yeast cells in the mutant pool. The amounts of barcode PCR produces, serve as a quantitative measure of the cell number of each deletion strain in the mutant pool. Traditionally, oligonucleotide microarrays have been used to deconvolute the identity of the strains in the mutant pool and quantify the amount of each barcode PCR product. In one of the recent studies, deep sequencing was found to perform equally well in this regard\cite{smith2009quantitative}. Barcode-based analyses of pooled mutants has been observed to contribute more significantly in improving the throughput of screens, reduce the amount of reagents used, and avoid the problems associated with strain cross-contamination than when compared to one-by-one screen of individual deletion mutants. The most frequently analyzed phenotype of pooled mutants is the growth rates, or fitness, of the mutant strains.From fitness
profiling of mutants under hundreds of growth conditions it has been observed that 97\% of the genes in the budding yeast genome are required for optimal growth under at least one condition \cite{han2010method}. In addition to phenotyping single-gene mutants, barcode-based analysis has also been used to study gene-gene interactions.

Barcodes have been utilized before as short genetic markers for multiple purposes such as identifying species \cite{hebert2004ten}, studying past diversity of the Earth's biota \cite{lambert2005large}. One of the  unique purpose where barcodes have been extensively used is knockout deletion of  genes from yeast strains. 
\subsubsection*{Process of barcode generation during Yeast gene knockout deletion }
 \DeclareGraphicsExtensions{.pdf,.png,.jpg}
\begin{figure}
\begin{centering}
    \includegraphics[width=1\textwidth]{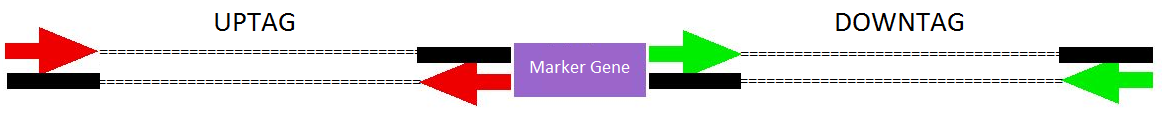}
    \caption{Barcode insert in a typical yeast genome \label{fig:1}}
      \end{centering}
\end{figure}
In the process of knockout deletion of  genes from yeast strains, each gene is systematically deleted  in the yeast genome. When completed, the result will be a number of mutant strains, each with a different gene deleted. However, rather than simply deleting the gene, a barcode is inserted which is a unique sequence of DNA for each strain. An analogy to barcodes can be made with the products packaging, used for checking out in retail shops, since a barcode is meant to be used as a unique identifier for each item. For the yeast barcode deletions, the unique identifier is composed of two unique DNA sequences called "up-tag" and "down-tag". Homologous recombination can be used to replace each gene with a unique barcode insert so that each strain had one gene deleted and replaced with its sequence-specific barcode insert. \ref{fig:1} shows a Prototype of a barcode insert used to replace genes in the yeast genome. The two arrows represent PCR primers, the series of N's represent the unique barcode sequences and the marker gene permits selection of cells that contain the insert. The up-tag consists of the 20 nucleotides on the left side and the down-tag consists of the 20 nucleotides on the right. The red PCR primer sequence would be the same in every barcode insert as would the green PCR primer sequence.

After each strain has its gene deleted, the challenge is to determine the functional consequences of all the deletions. For doing that in a high throughput method rather than each strain individually mutants are screened efficiently. About 500 mutants are put into a common flask and two PCR primers is used to amplify all 500 barcode inserts from a mixture of genomic DNA extracted from an aliquot of the 500 strains\cite{YEASTBARCODE}. These PCR products are tagged with a red dye. The cells were allowed to grow over an extended period of time. During this time, strains with deleted genes needed to maintain normal growth rate would constitute a reduced percentage in the growing population. At various times, aliquots of cells are removed and the genomic DNA is extracted and amplified using the same two PCR primers. The PCR products used in the later condition are tagged with a green dye. Genomic DNA from an aliquot of this initial population of cells is used as template for PCR using the two primer sequences shared by each of the deletion strains. These PCR products are labeled red. After a period of time, an aliquot of cells are again used for template of PCR using the same two PCR primers. These PCR products are colored green.


\DeclareGraphicsExtensions{.pdf,.png,.jpg}

\begin{figure}[h!]
\small
\begin{centering}
    \centerline{\includegraphics[width=0.5\textwidth]{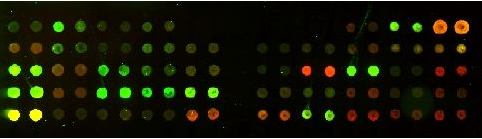}}
    \caption{Three possible labelling patterns for its associated up-tag and down-tag \label{fig:3}}
      \end{centering}
 \end{figure}

Barcode yeast experiments require a unique DNA microarray. Instead of spotting the coding DNA for each gene, the uptag and downtag for each deletion was spotted onto the glass. The two PCR products (red from time zero and green from a later time) are mixed together and incubated with the DNA microarray. For each strain there could be three possible lableing patterns for its associated uptag and downtag given in figure \ref{fig:3}.

From figure \ref{fig:3} one can make the following conclusions
\begin{itemize}
\item
 A strain growing a yellow spot indicates that its proportion of the initial and final total population was the same.
\item
A strain growing a red spot indicates that the strain grew slower than most other cells. Hence, its uptag and downtag spots showed as red since this strain was present at a higher concentration in the initial population than in the later population.
\item
A strain growing a green spot indicates that the strain grew faster than most other cells. Hence, its up-tag and down-tag spots appeared more green than yellow since its increased rate of growth resulted in an increased representation of the total population.
\end{itemize}

Figure \ref{fig:3} shows DNA microarray showing yellow, red and green spots for cells that maintained the most common growth rate, decreased growth rate or increased growth rate, respectively. If a strain's growth rate were truly affected by the deletion, then both the uptag and downtag spots should exhibit similar results.

\subsubsection*{Bias incorporated during PCR amplification }

A major drawback which all the PCR based NGS systems share is the amplification bias. It has been observed that barcode-dependent bias is highly prominent in DNA and RNA detection and expression profiling using multiplexing technology  \cite{alon2011barcoding,berry2011barcoded }. For applications where many cycles of PCR are required for sensitive detection, bias and noise reduction are crucial for accurate quantiﬁcation. In one of the recent studies \cite{van2011quantitative}, an in-depth quantitative measurement of PCR amplification bias, resulting from the use of barcodes has been conducted. This research attempted to adapt previous barcoding strategies to multiplexed sequencing of small RNA used index sequences placed at the distal end of the 5’ adapter in the Illumina small RNA library protocol. The observations made in that study was that despite a number of iterations of the design there was a consistent failure in avoiding PCR amplification bias when identical samples with different barcodes were compared. Though in this study the authors showed that introducing the barcode during the PCR instead of before the PCR step using differentially, barcoded primers does not result in bias, it remained unclear as to It is unclear as to why pre-PCR protocol for small RNA produces biased results, while the TruSeq protocol for mRNA/dsDNA produces unbiased \cite{van2011quantitative} results.



\section*{Methods}

\subsection*{Dataset used}
Data Set used is SGA Barseq data for Yeast strains. The data contain 3880 unique yeast strains with 3880 unique barecodes. Total number of features generated from the barcodes is 275 which are considered as the predictor variables.  The features are given below are all possible Single nucleotide frequencies, double nucleotide frequencies, Triple nucleotide frequencies, melting point of each nucleotides, longest 2-mers of repetition of A, C, T and G, longest 3-mers of repetition of A, C, T and G, repetition of AT, repetition of CG, frequency of last 3 base pairs of the barcode sequence and first three base pair of the sequence (see figure \ref{fig:15}). Response variable is the read count which has been considered for classification and prediction purpose. All the predictor variables are used to detect the response pattern to see which feature / predictor variable causes reads of high count. Motif discovery has been performed over the original barcode sequence features. The input data for Motif discovery is the yeast strain Identifier profile and the expression profile which is the read counts in this project.

\subsection*{Proposed approach}

The purpose of this project is to detect the probable causes behind PCR amplification bias which is observed in Barcode sequences used for DNA analysis of Yeast. This analysis is entirely based on the sequence features of the barcodes used and does not considers any kind of contribution in the bias from any steps of ligation or PCR. Our target in this project is two fold - 
\begin{itemize}
\item
Building a prediction model using all possible features of the barcode sequences and analyze if this model can bring about a prediction in the read count.
\item
Important motif discovery in the barcode sequences based on Mutual information \cite{elemento2007universal} and analysing their probable contribution in the prediction model to detect the cause of the bias.
\end{itemize}  

The prediction technique used is regression analysis - Lasso regularization method which aims at reducing the number of redundant predictors and identifying important predictors from the set of all features generated from the bar code sequences. Another important technique has been used is the motif discovery. With the help of this technique important motifs and their pattern of occurrence can be discovered. The classification and prediction technique used in this paper is the lasso regression technique and mutual information based motif discovery from barcode sequences. Lasso is a regularization technique for performing linear regression. The main motive behind using Lasso is that 
\begin{itemize}
\item
Lasso reduces the number of predictors in a regression model. In our data this will help eliminating those variables/features of barseq which produces minimum or no significant contribution for the prediction of read counts.
\item
Lasso identifies important predictors and select among redundant predictors. This will allow us concentrate upon sequence feature which causes an increased read counts. 
\item
Produce shrinkage estimates with potentially lower predictive errors than ordinary least squares.
\end{itemize}

 Lasso includes a penalty term that constrains the size of the estimated coefficients. Lasso can also be called a shrinkage estimator since it generates coefficient estimates that are biased to be small. Nevertheless, a lasso estimator can have smaller mean squared error than an ordinary least-squares estimator when you apply it to new data.For a given value of lambda, a nonnegative parameter, lasso solves the problem given in equation \ref{fig:5}. where $N$ is the number of observations, $y_i$ is the response at observation $i$, $x_i$ is data, a vector of $p$ values at observation $i$, $\lambda$ is a positive regularization parameter corresponding to one value of $\lambda$, The parameters $\beta_0$ and $\beta$ are scalar and vector respectively.

 \begin{equation}	\label{fig:5}	
		\min_{\beta_0, \beta} \left(\frac{1}{2N}\sum_{i=1}^{N}\left( y_i - \beta_0 - x_i^T\beta \right)^2 + \lambda \sum_{j=1}^{p} \mid \beta_j\mid \right) 
 \end{equation}

A 10 fold cross validation is used while using lasso to eliminate overfitting of the model with the data used. Manual cross validation has also been used where data is divided into 7:3 ratio for prediction of read counts. 

Motif discovery has been performed over the original barcode sequences in the past researches. This approach infers motifs from gene expression data that aims at making as few a prior assumptions as possible by quantifying the dependency between the presence or absence of a given motif in a regulatory region and the expression of the corresponding gene \cite{elemento2007universal}. The concept of mutual information is used for discovering motifs whose patterns of presence/absence across all considered regulatory regions are most informative about the expression of the corresponding genes. Thus, knowing whether such a motif is present or absent within the regulatory region of a given gene provides significant information regarding the expression of that gene which is in our case the read counts. By discovering important motifs from the barcode sequences it can be possible to predict which motif characters are responsible for causing a bias during amplification. The mathematical formulation of mutual information is given in equation \ref{fig:5} where $p(i,j)$ is the joint probability distribution function of the random variables $i$ and $j$, and  $p(i)$ and $p(j)$ are the marginal probability distribution functions of i and j respectively.

 \begin{equation} \label{fig:6}
 	I(motif;expression) = \sum_{i=1}^2\sum_{j=1}^{N_e}P(i,j)log\frac{P(i,j)}{P(i)P(j)} 
 \end{equation}

A detailed description of the results obtained from the above two approach is given in the result section.


\begin{figure}
\small
\begin{centering}
   \centerline{\includegraphics[width=0.6\textwidth]{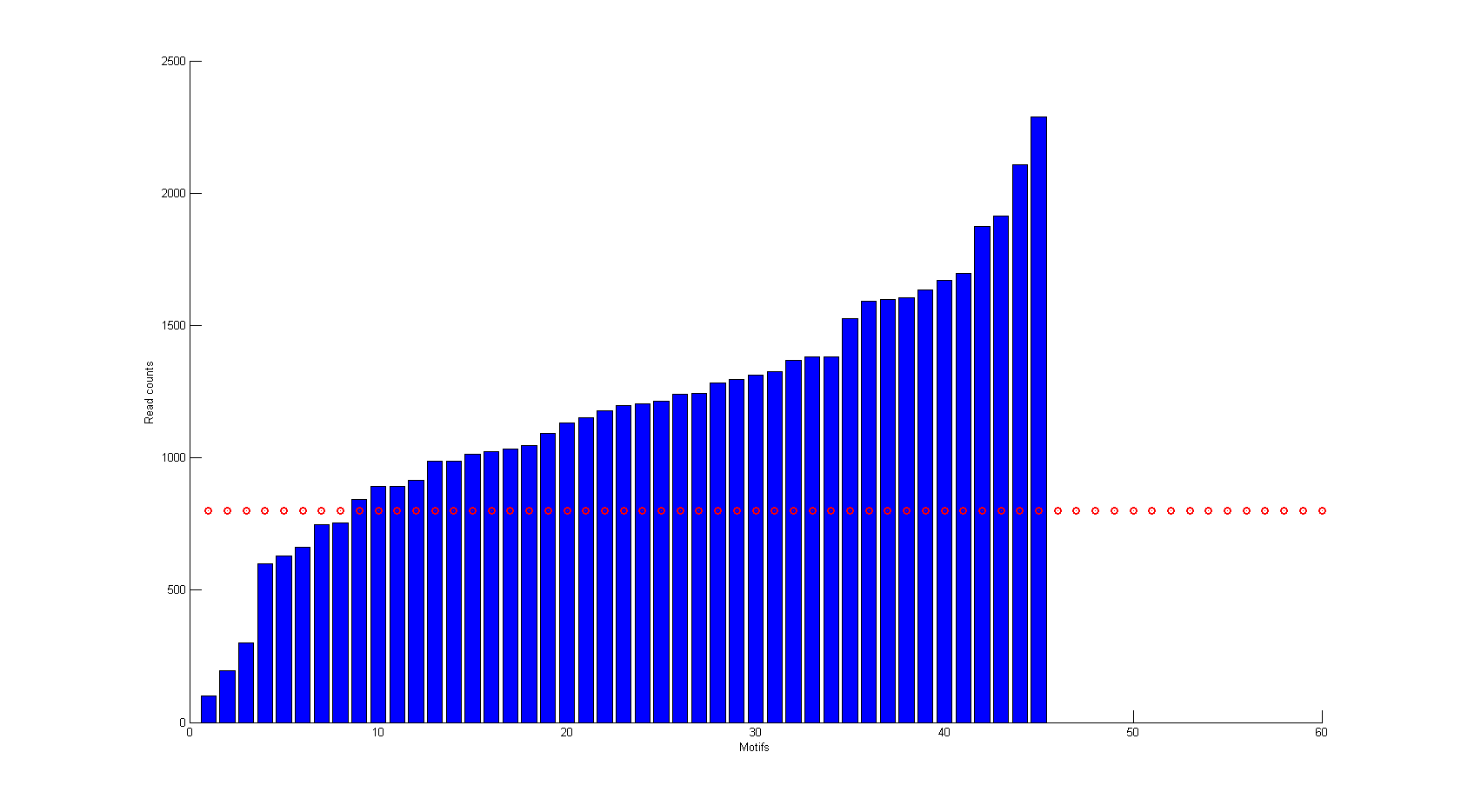}}
    \caption{Average read counts. Red dots shows the boundary for reads greater than 800. It can be seen that approximately 80 \% of sequence having the average read counts greater than 800   \label{fig:7}}
      \end{centering}
 \end{figure}

\begin{figure}
\small
\begin{centering}
    \centerline{\includegraphics[width=0.4\textwidth]{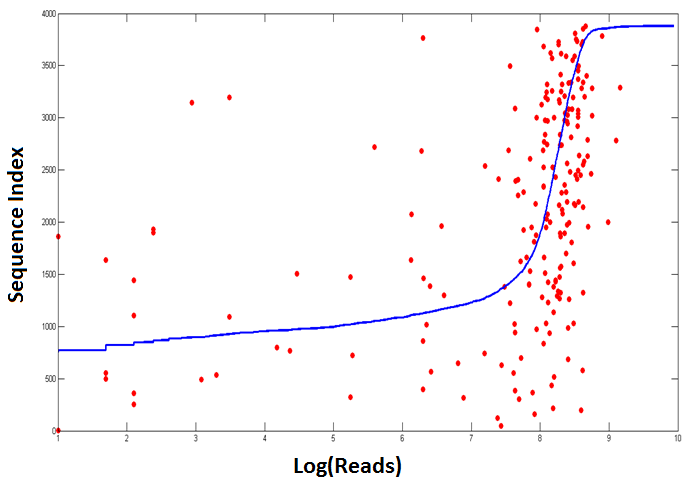}}
   \caption{Actual sequence reads plotted along with the scatter plot of sequence indexes containing all 3,4,5,6,,7 and 8 mer motifs  \label{fig:8} }
      \end{centering}
 \end{figure}

\begin{figure}
\small
\begin{centering}
 \centerline{\includegraphics[width=0.5\textwidth]{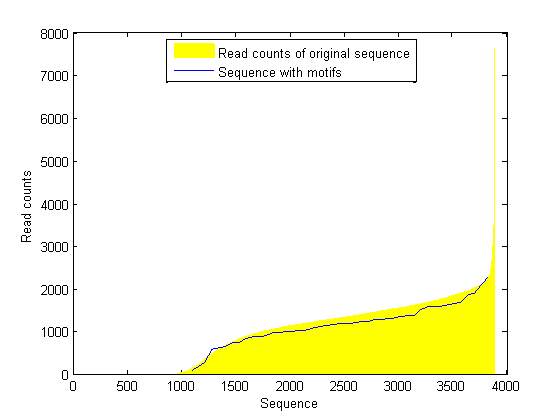}}
    \caption{Distribution of reads with sequence containing motifs compared with distribution of the reads of all sequence \label{fig:9}}
      \end{centering}
 \end{figure}
 
 \begin{figure}
\small
\begin{centering}
    \centerline{\includegraphics[width=.8\textwidth]{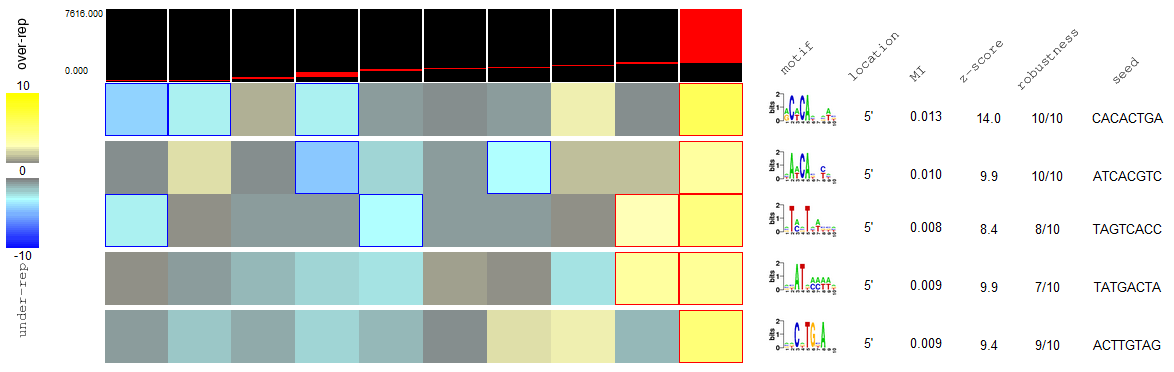}}
    \caption{8 -mer summery of the motif discovery tool FIRE showing under and over representation of the motifs in the sequences based on the reads 
  \label{fig:10}}
      \end{centering}
 \end{figure}

 \begin{figure}
\small
\begin{centering}
  \centerline{\includegraphics[width=0.8\textwidth]{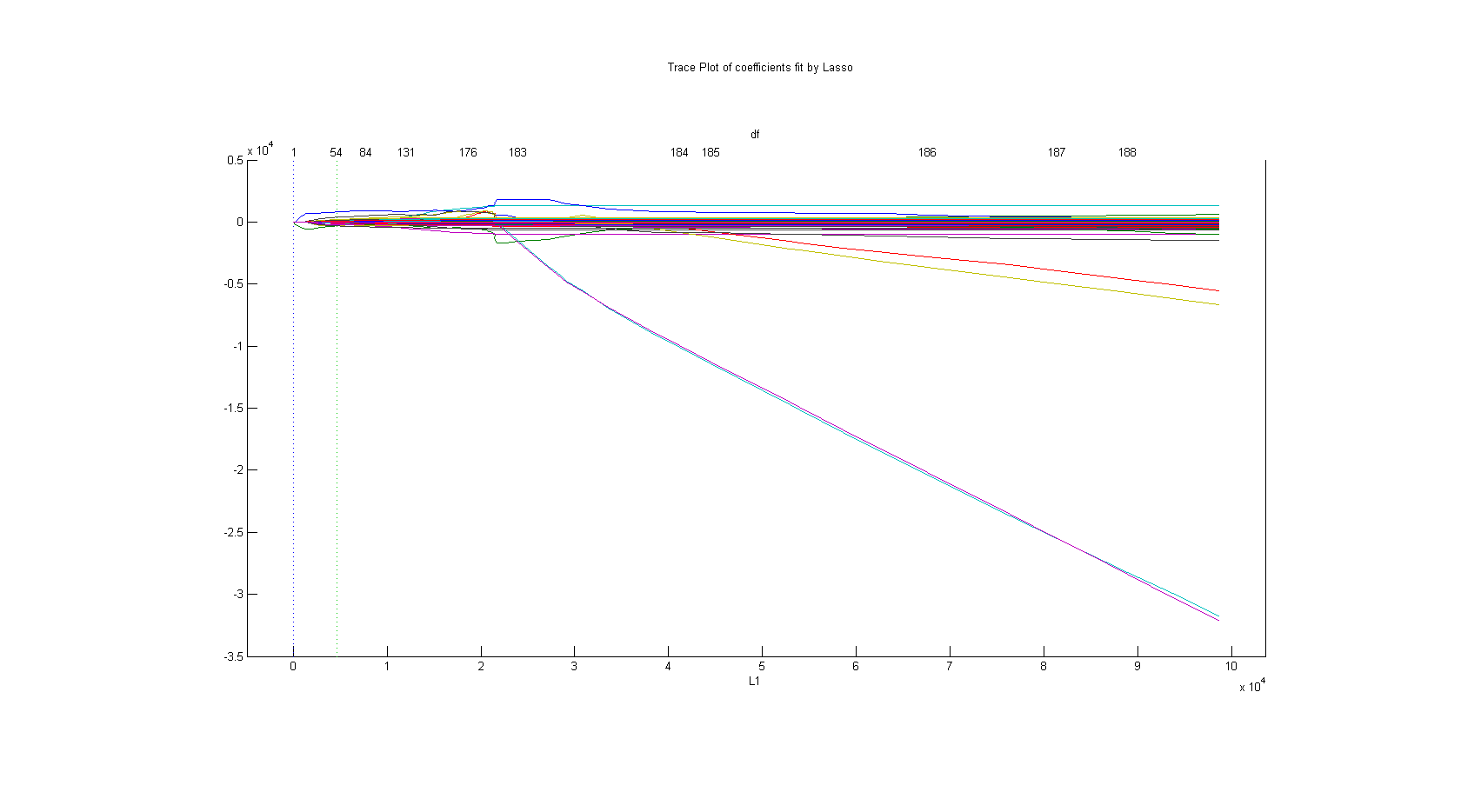}}
   \caption{Lasso Regression Plot with 229 features  
  \label{fig:11}
   }
      \end{centering}
 \end{figure}
 
 \begin{figure}
\small
\begin{centering}
  \centerline{\includegraphics[width=0.5\textwidth]{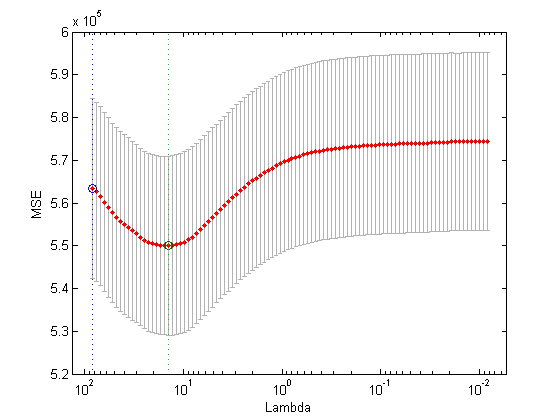}}
   \caption{MSE vs Lambda in Lasso Regression with 299 features     \label{fig:12}
  }
      \end{centering}
 \end{figure}

 \begin{figure}
\small
\begin{centering}
    \centerline{\includegraphics[width=0.8\textwidth]{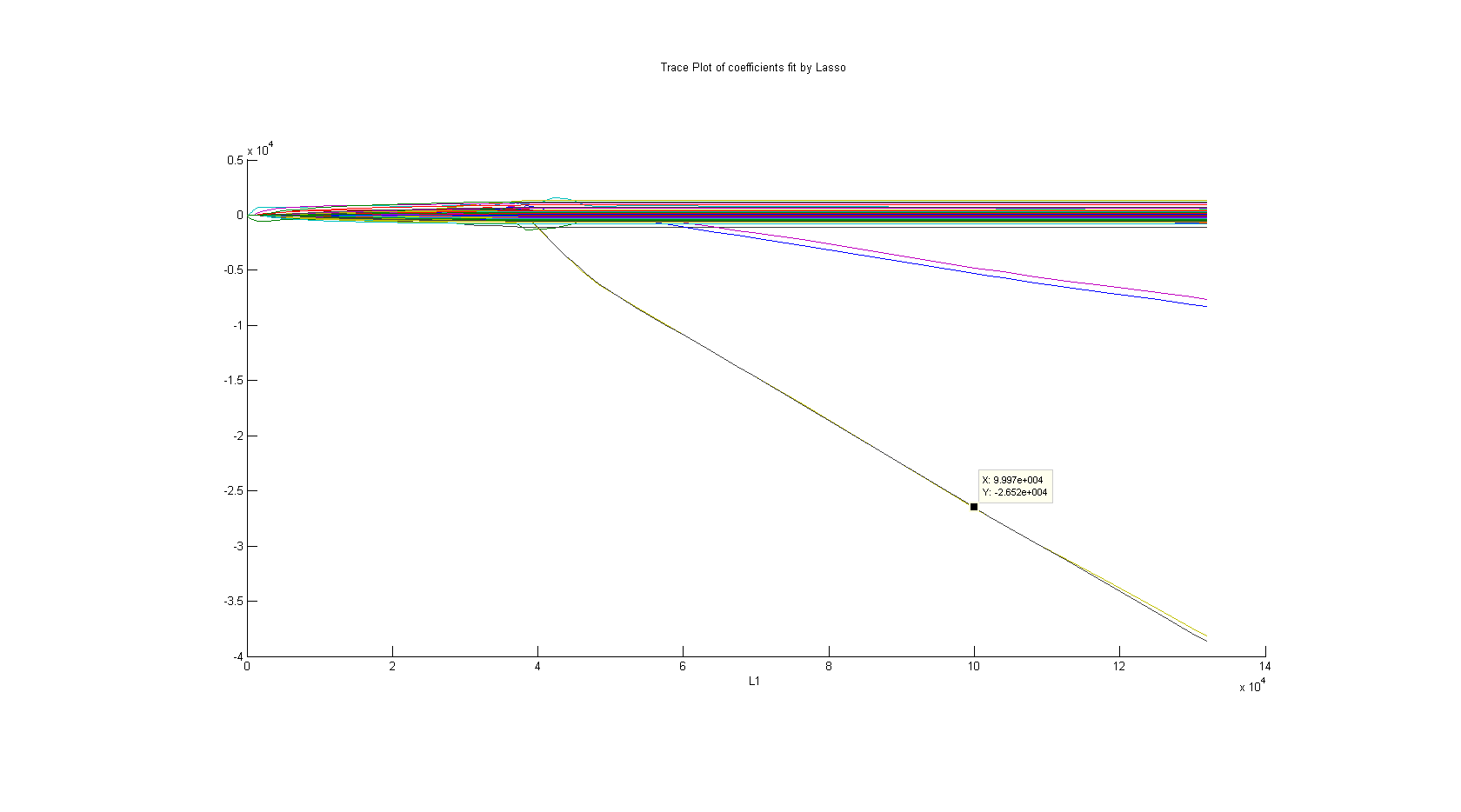}}
   \caption{Lasso Regression Plot with 275 features  \label{fig:13}}
      \end{centering}
 \end{figure}
 
 \begin{figure}[h!]
\small
\begin{centering}
   \centerline{\includegraphics[width=0.5\textwidth]{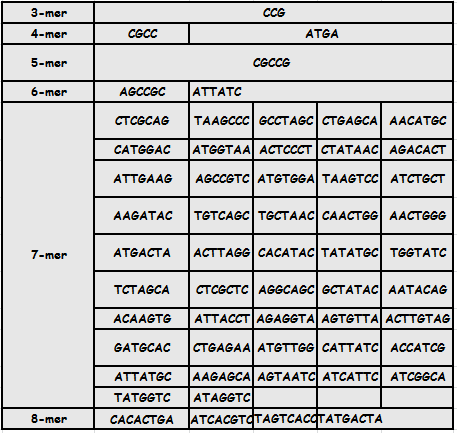}}
   \caption{k-mer Motifs obtained using mutual information based profiling \label{fig:14}}
      \end{centering}
 \end{figure}

\section*{Results}
First the barcode sequences were analyzed using FIRE tool \cite{elemento2007universal}. In this approach each barcode sequence bases  are used for finding seed motifs in order to generate motif profile. Then an expression profile is created using the read counts for all the 3880 barcode sequences. The FIRE algorithm uses the motif profile and the expression profile to generate mutual information for each motif. Then the list of motifs discovered is sorted and sort listed for choosing those motifs which has highest mutual information. In other words, those motifs are selected whose  profiles are highly informative about the behavior profile which is the read counts.The result is a list of motifs that are generally over- or under-represented in the sequences associated with certain behavior categories/bins.  The motifs that we discovered from the barcode sequences are shown in figure \ref{fig:14}

From all the k-mers that we obtained where k is 3,4,5,6,7 and 8 80\% motifs discovered were 7-mers. The motifs discovered are diverse in their A,C,G,T features but a frequent occurrence of CCG , GCC , TAT, ATAT, has been observed. Most of the Motifs either starts with A or C. However, the significance of such patterns can be analyzed only after a thorough examination of the motifs. In figure \ref{fig:10} a snap shot of 8-mer motif profile as obtained from FIRE has been provided. This figure shows the yellow regions which represents over-representation of the motifs corresponding to a higher read count (read box on top right). Over representation of these motifs corresponds to active motifs i.e. represent putative functional instances and under representation denotes motifs that are more likely to be non-functional.

After discovering the motifs with the highest mutual information we analyzed them with our actual barcode sequence. We parsed the sequences in order to record the possibility of occurrences of the motifs in the barcodes and the read counts associated with the original barcode sequences. Figure \ref{fig:7} represents a bar graph with the average read measure for all the sequences that contained the motifs. From figure \ref{fig:7} we can see that approximately 80\% of the sequences containing motifs have more than 800 average read counts which is much higher as compared with low read counts considering the range of 10 - 2500. Also in figure \ref{fig:8} we have plotted the actual reads of all the sequences vs the scatter plot (red dots) of the sequences which contained the motifs. We can observe a dense red dotted area near the region of high reads. This may indicate that sequence containing the motifs have a tendency for higher read counts.

In figure \ref{fig:9} we plotted the distribution of the reads corresponding to all the sequences to compare it with the distribution of the sequences containing motifs. This figure depicts an interesting outcome where we see that the 46 sequences have the same distribution as that of the entire population i.e. the entire barcode sequences. This can indicate the fact that though most of the motifs shows a higher read count , some motifs are present with lowers the read counts as well. These outliers are some thing which can be worth investigating.

\begin{figure}[h!]
\small
\begin{centering}
   \centerline{\includegraphics[width=0.5\textwidth]{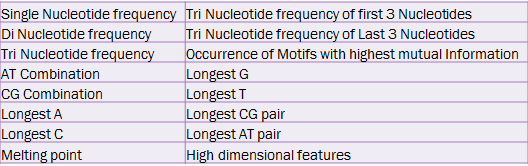}}
   \caption{Sequence features generated from the barcode sequences \label{fig:15}}
   \end{centering}
 \end{figure}

After the motif discovery, we generated sequence features based on A,C,G,T base pairs of the barcode sequences. The features that we generated are given in figure \ref{fig:15}. We generated 275 features and applied Lasso Regression technique for predicting the features which has the maximum contribution in classifying the read counts. At first we applied this prediction algorithm over 229 raw features that had been derived from the barcode sequences. The results of the lasso plot of L1 norm vs the penalty term lambda (given in figure \ref{fig:11}) and lasso plot of mean square error(MSE) vs lambda (given in figure \ref{fig:12})  shows that it was able to predict the best value of lambda which gave approximately 15 features that contribute most in prediction with the minimum MSE. The over all correlation between the actual reads and the predicted reads we obtained, from different values of lambda ranges from 0.16 to 0.28. Our next step was to incorporate the motifs as additional features along with the previously analyzed 229 features using lasso technique. We added 46 more predictors to our sequence feature data. This value was binary. If a feature contains the motif the data would be 1 and if not 0. The result of lasso over 275 features is shown in Figure \ref{fig:13}. The correlation between the predicted reads and the actual reads was obtained - which range  from 0.16 to 0.3410. This correlation measure is slightly better than the previously obtained result where motifs were not used as the predictors. 
These results were obtained from analyzing the barcode features using the two methodologies mentioned in the previous section. Further analysis is done in the Discussion section.

\section*{Discussion}

The Next-generation sequencing have revolutionized modern genomics as well as have increased our understanding about DNAs and RNAs in particular \cite{alon2011barcoding}. However, this still-growing field inevitably creates some biases in the vast amount of data generated. Analyzing the reason behind these biases is a crucial task especially when using multiplexing techniques with barcodes for accuracy and sensitivity. The main aim of our project has been an attempt in solving this riddle. Though a lot of trial and error with PCR cycles and ligation procedure has been done previously to reduce this bias, but no studies have been conducted to the best of our knowledge which have analyzed barcode sequences. We have performed a classification prediction based technique  as well as mutual information based motif discovery technique in order to analyze the barcode features. This analysis shows that there is correlation existing between the barcode features and the read count. Most importantly we could find the motifs or sub-sequences that contributed to that correlation. It clearly proves that these motifs play an important role in determining the read counts.  Also we could analyze those sequence features which did not have any influence on predicting the read counts. This is also a clear indication that the motifs discovered are important since the correlation between the predicted reads and the actual reads increased when the motifs were used as additional predictors. However, the actual pattern in which these motifs behave could not be analyzed due to time constraint of this project. Given more time it is not impossible to generalize the motif patterns which might predict the way the read counts behave.

 \bibliographystyle{amsplain}
\bibliography{BarCodeSequencing}

\end{document}